# Realizing high current gain PNP transistors using a novel Surface Accumulation Layer Transistor (SALTran) concept

M. Jagadesh Kumar[1] and Vinod Parihar

Department of Electrical Engineering,

Indian Institute of Technology Delhi,

Hauz Khas, New Delhi – 110 016, INDIA.

Email: mamidala@ieee.org Fax: 91-11-2658 1264

[1]Corresponding author

**Abstract:** In this paper we report a new PNP **S**urface **A**ccumulation **L**ayer **Tran**sistor (SALTran) on SOI which uses the concept of surface accumulation of holes near the emitter contact to significantly improve the current gain. Using two-dimensional simulation, we have evaluated the performance of the proposed device in detail by comparing its characteristics with those of the previously published conventional PNP lateral bipolar transistor (LBT) structure. From our simulation results it is observed that depending on the choice of the emitter doping and the emitter length, the proposed SALTran exhibits a current gain enhancement of around 20 times that of the compatible lateral bipolar transistor without deteriorating the cut-off frequency. We have discussed the reasons for the improved performance of the SALTran based on our detailed simulation results.

# 1. Introduction

Complementary bipolar technologies using NPN and PNP transistors play an important role in many analog applications such as feedback amplifiers, current mirrors and push-pull circuits [1]. The presence of PNP BJTs in the output stage is crucial for improved driver performance. Also, active loads in analog applications can not be implemented without PNP transistors. However, realizing compatible PNP and NPN transistors is difficult because PNP transistors exhibit low current gains due to poor hole mobility. Even the use of SiGe base [2] and polysilicon emitter [3] could not be of much help in realizing large gain PNP transistors. While both the techniques are widely used, they require complex process steps. Another technique which has been reported to increase the current gain of a bipolar transistor is the application of low-high emitter junction[4-7]. Nonetheless, in this case, the cut-off frequency deteriorates due to the increase of minority carrier transit time caused by the presence of the high-low junction[7]. While high gain lateral PNP transistors can be realized in submicron CMOS technology using complex processes, it would be of great practical importance, if the current gain of a PNP BJT can be enhanced using a simple emitter contact concept while obviating the above difficulties.

The aim of this work is therefore to present for the first time a new high current gain PNP bipolar transistor using a **S**urface **A**ccumulation **L**ayer **Tran**sistor(or SALTtran) concept. The process steps in the SALTran are similar to that of a conventional bipolar transistor except that in the SALTran a lightly doped emitter with a metal contact whose workfunction is higher than that of the silicon is employed. This results in an accumulation of holes at the emitter contact forming a reflecting boundary for the electrons injected into the emitter from the base region leading to a significant reduction in the base current. We demonstrate using two-dimensional simulation that the proposed SALTran is superior in performance compared to an equivalent lateral PNP BJT on SOI in terms of high current gain. We further demonstrate that unlike in the case of low-high emitter bipolar transistors, the cut-off frequency of the SALTran does not degrade by the presence of the reflecting boundary in the emitter if optimized emitter doping and length are used.

## 2. The Surface Accumulation Layer Transistor(SALTran) Concept

The SALTran is based on a concept that when a metal of high work function is brought in contact with a lightly doped p-type semiconductor such that the metal's workfunction is higher than that of the semiconductor, an accumulation of holes takes place in the semiconductor near the metal-semiconductor interface[8]. This results in an electric field due to the hole concentration gradient from the metal-semiconductor interface towards the emitter-base junction. The direction of this field is such that it causes the reflection of electrons injected from the base into the emitter resulting in a reduced electron concentration gradient in the emitter region. Thus the application of such a reflecting boundary contact to the emitter region should result in a reduced base current leading to a significant improvement in the current gain.

In the following sections, we shall demonstrate using accurate two dimensional simulations that the surface accumulation layer emitter contact does indeed enhance the PNP bipolar transistor performance significantly.

## 3. Design Methodology of SALTran

To demonstrate the concept of the SALTran on SOI and to calibrate our device simulations, we have first chosen the typical experimental process steps for an SOI PNP lateral bipolar transistor (LBT) reported in literature[9]. We have implemented the fabrication steps of [9] in the process simulator ATHENA[10]. The top layout and the schematic cross-section of the SALTran or the LBT are shown in Fig.1. The final doping profiles of the SALTran and the LBT obtained using ATHENA are shown in Fig. 2. The simulation parameters are given in Table 1. We have imported the structure and doping profiles into the device simulator ATLAS[11] and calibrated the default model parameters such that the simulated current gain of the lateral SOI BJT matches with the reported values in [9]. We have adjusted the emitter implantation parameters in ATHENA to obtain different emitter doping values between $1\text{x}10^{14}$ $cm^{-3}$ and $1\text{x}10^{18}$ $cm^{-3}$ for the SALTran and evaluates the electrical characteristics of both the structures as discussed in the following section.

## 4. Simulation Results and Discussion

### *A) Device Characteristics*

In our simulations using device simulator ATLAS, we have used suitable models for the bandgap narrowing, SRH and Auger recombination and the concentration and field dependent mobility. The simulated output characteristics of the SALTran is shown in Fig. 3 for an emitter doping of $1\text{x}10^{14}$ $cm^{-3}$ and emitter length of 0.3 μm. The simulated energy band diagram for the SLATran and the LBT without and with bias are shown in Fig.4 which clearly shows the band bending at the metal-semiconductor junction indicating the formation of hole accumulation. The Gummel plot of the SALTran shown in Fig. 5(a) indicates that the base current in the SALTran is quite smaller than that of the LBT resulting in an enhanced current gain as shown in Fig. 5(b). We further notice that with the reduction in emitter doping from $10^{18}$ $cm^{-3}$ to $10^{14}$ $cm^{-3}$, the base current decreases significantly resulting in a drastic improvement in the current gain. As we changed the emitter length from 3.8 μm to 0.3 μm, the peak current gain of the LBT decreased from 14 to 2, while that of the SALTran increased from 120 to 235. The reduction in the current gain of the LBT for smaller emitter lengths is due to the reduction in the emitter Gummel number. The significantly high current gain in case of the SALTran can be understood from the hole profile and the electric field profile shown in Fig. 6. As pointed out in Section 2, since we have chosen the workfunction(5.4 eV) of the emitter titanium metal contact to be greater than that of the silicon emitter region, there is an accumulation of holes under the metal contact as shown by the simulated hole profile in Fig. 6(a) for the SALTran structure with two different dopings. No such hole accumulation is observed in the case of the LBT. The accumulated hole gradient in the SALTran results in a large electric field under the emitter as shown in Fig. 6(b) and acts as a reflecting boundary for the electrons arriving from the emitter-base junction. The base current of the SALTran will, therefore, be significantly smaller than that of the LBT as demonstrated in the Gummel plots of Fig. 5.

The simulated cut-off frequency $f_T$ of the SALTran and the LBT are compared in Fig. 7. The cut-off frequency of the SALTran is almost same as that of the LBT. This is unlike the behaviour shown by the high-low junction emitter bipolar transistors in which while a gain enhancement is observed due to the presence of the high-low junction but the cut-off frequency deteriorates due to an increase in the emitter transit time because of charge storage effects[7].

***B) Emitter Region Optimization***

An important aspect of the SALTran structure is that as the emitter length decreases both the current gain and cutoff frequency begin to increase. A reduction in the emitter length causes the rate of reflection of the electrons from the emitter to be significant. Fig. 8 shows the peak current gain variation for different emitter lengths and emitter dopings for both the SALTran and the LBT structures. We notice that when the doping is far smaller than the peak base doping, the current gain enhancement is maximum. For example, when the emitter doping is $1 \times 10^{14}$ $cm^{-3}$, the current gain enhancement is largest for shallow emitter depths which makes it very attractive for scaled down VLSI BJTs. We also notice that for deeper emitter depths too the current gain enhancement is still impressive when the emitter is lightly doped. This is an indication that the SALTran concept can be conveniently applied to even high voltage PNP power bipolar transistors in which low current gain is often a problem. In the case of deeper emitter junctions, the current gain of the SALTran is smaller than that of shallower emitter due to the increased emitter region recombination for deeper emitters.

The peak cutoff frequency variation with emitter length and emitter doping is shown in Fig. 9. It is seen that with proper optimization of emitter length and emitter doping we can realize a high current gain and a high cutoff frequency.

# 5. Conclusions

For the first time, the concept of surface hole accumulation to increase the current gain and also the cutoff frequency of PNP bipolar transistors is successfully shown by using 2-D simulation. We have demonstrated that the presence of a reflecting boundary for the minority carriers at the emitter contact results in a significant improvement in the current gain of the bipolar transistors. Unlike in the case of the high-low emitter bipolar transistors, the cut-off frequency of the SALTran does not deteriorate in the presence of surface hole accumulation if optimized emitter design is used. Since the SALTran structure obviates the need to create a high-low junction in the emitter region and since its performance improves for both shallow and deep emitters, the proposed PNP SALTran concept should be useful to the designers to enhance the bipolar transistor performance in analog VLSI applications where high performance PNP transistors are often required.

Table 1: ATLAS input parameters used in the simulation of SALTran/LBT

| Parameter | Value |
|---|---|
| Silicon film thickness $t_{si}$ | 0.20 μm |
| Buried oxide thickness $t_{box}$ | 0.38 μm |
| Metal work function for emitter contact(Titanium) | 5.4 eV |
| Emitter length | 0.3 – 3.8 μm |
| Base length | 0.40 μm |
| Collector length | 3.8 μm |
| Emitter region doping level | $1x10^{14}$ $cm^{-3}$ to $1x10^{18}$ $cm^{-3}$ |
| Base region doping level | $5x10^{17}$ $cm^{-3}$ |
| Collector region doping level | $1x10^{14}$ $cm^{-3}$ |
| $P^+$ collector region doping level | $5\ x10^{19}$ $cm^{-3}$ |
| SRH electron minority carrier life time coefficient (TAUN0) | $6.5x10^{-6}$ s |
| SRH hole minority carrier life time coefficient (TAUP0) | $1x10^{-6}$ s |
| SRH concentration parameter for electrons and holes NSRHN and NSRHP | $5x10^{16}$ $cm^{-3}$ |
| Surface recombination velocity at poly base contact (VSURFP) | Atlas' default |

## Figure captions

.

Fig. 1 Top layout and schematic cross-section of the SALTran and the LBT structures.

Fig. 2 Typical doping profile of the SALTran and the LBT structures.

Fig. 3 Output characteristics of (a) the SALTran, (b) the LBT structures.

Fig.4 Energy band diagram of the SALTran and the LBT (a) without bias and (b) with bias.

Fig. 5 (a) Gummel plots and (b) Current gain versus collector current of the SALTran and the LBT structures.

Fig. 6 (a) The hole concentration and (b) the Electric field of the SALTran and the LBT structures in the emitter.

Fig. 7 Cutoff frequency versus collector current for the SALTran and the LBT structures.

Fig. 8 Current gain versus emitter length of the SALTran for different emitter dopings ($1x10^{14}$ /$cm^3$ to $1x10^{18}$ /$cm^3$ ) and the LBT structure for an emitter doping of $5x10^{19}$ /$cm^3$.

Fig. 9 Cutoff frequency versus emitter length of the SALTran for different emitter dopings ($1x10^{14}$ /$cm^3$ to $1x10^{18}$ /$cm^3$ ) and the LBT structure for an emitter doping of $5x10^{19}$ /$cm^3$.

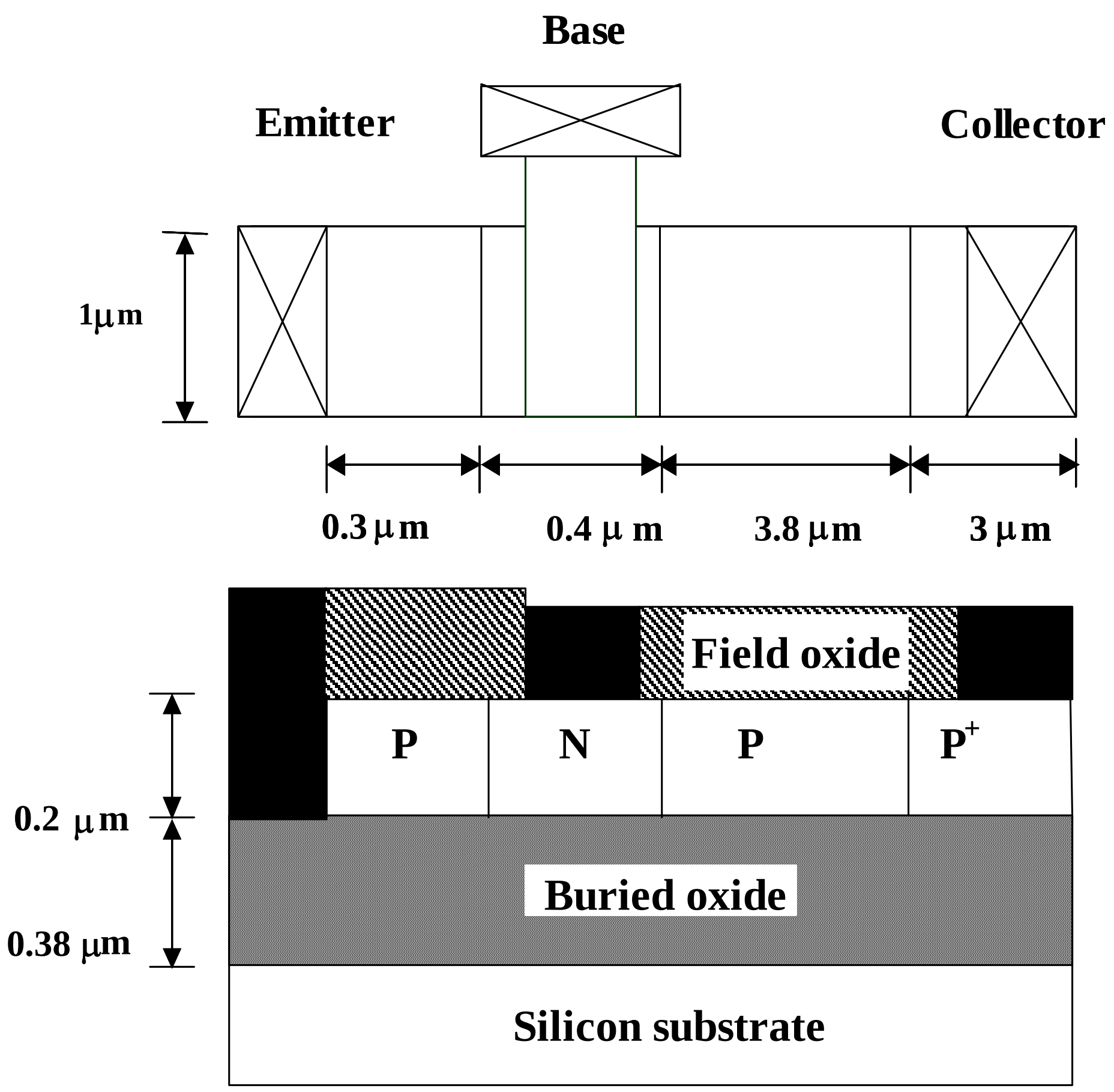

Fig. 1

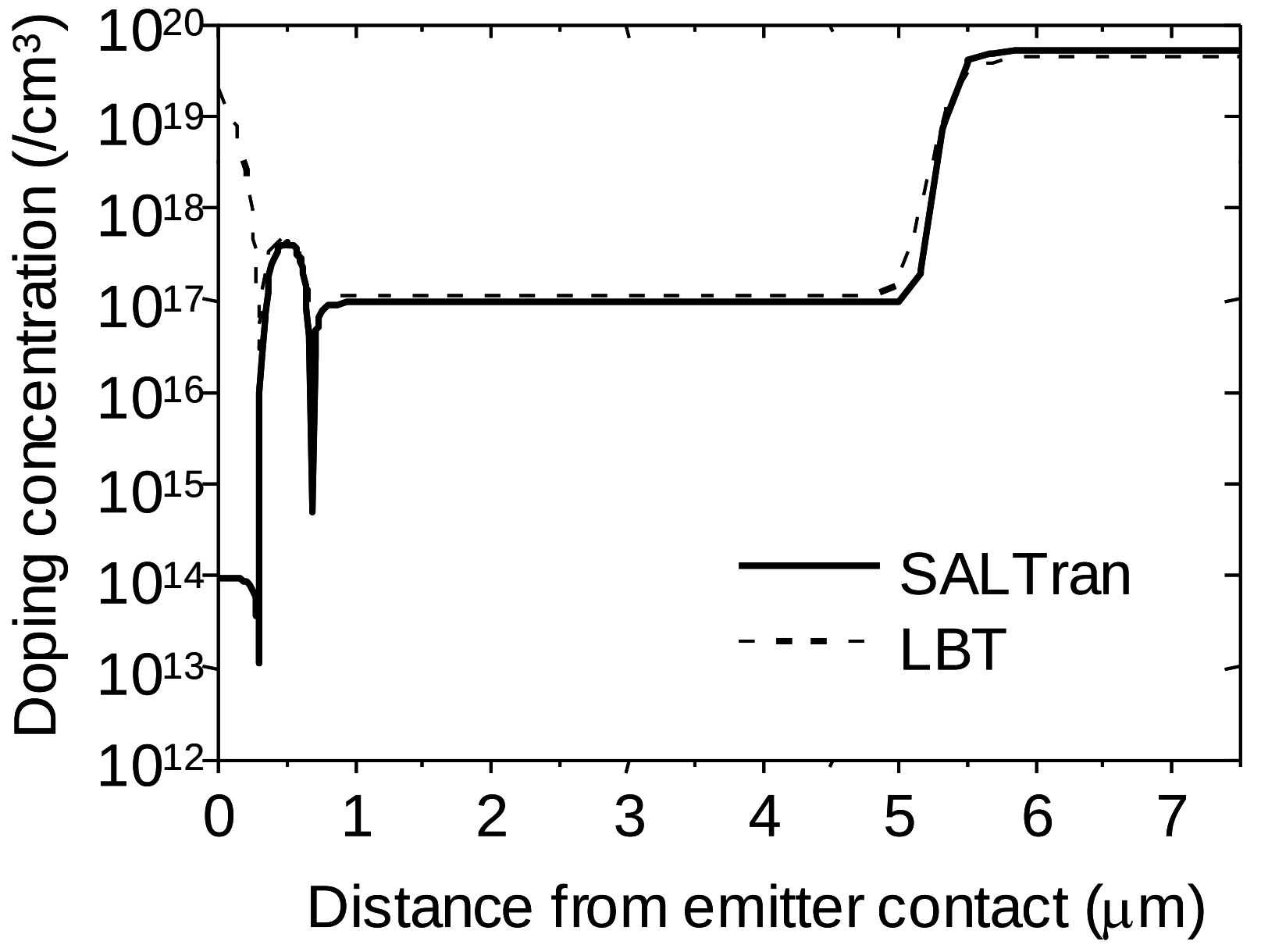

Fig. 2

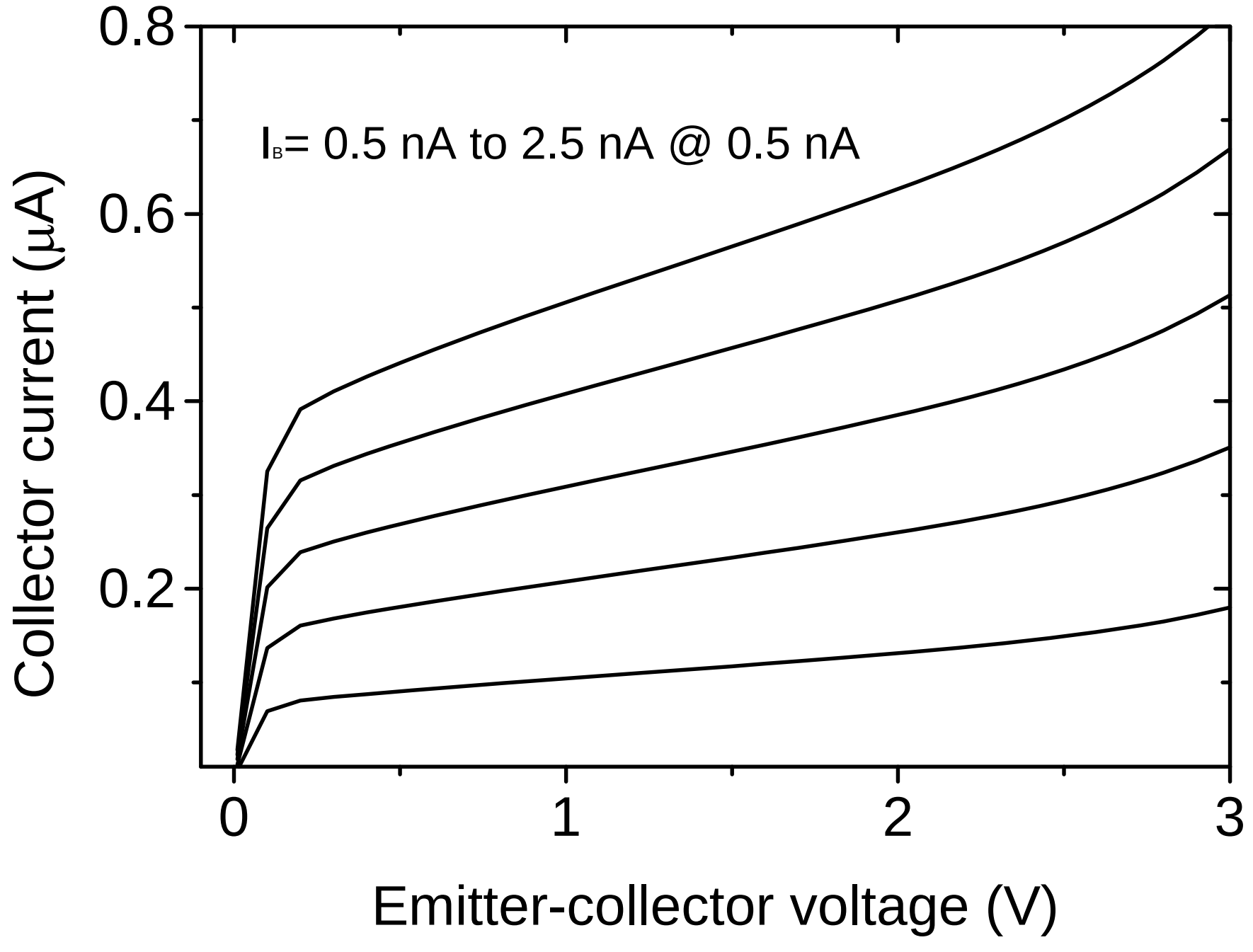

Collector current (μA)
0.8
0.6
0.4
0.2
IB= 0.5 nA to 2.5 nA @ 0.5 nA
0
1
2
3
Emitter-collector voltage (V)

Fig. 3

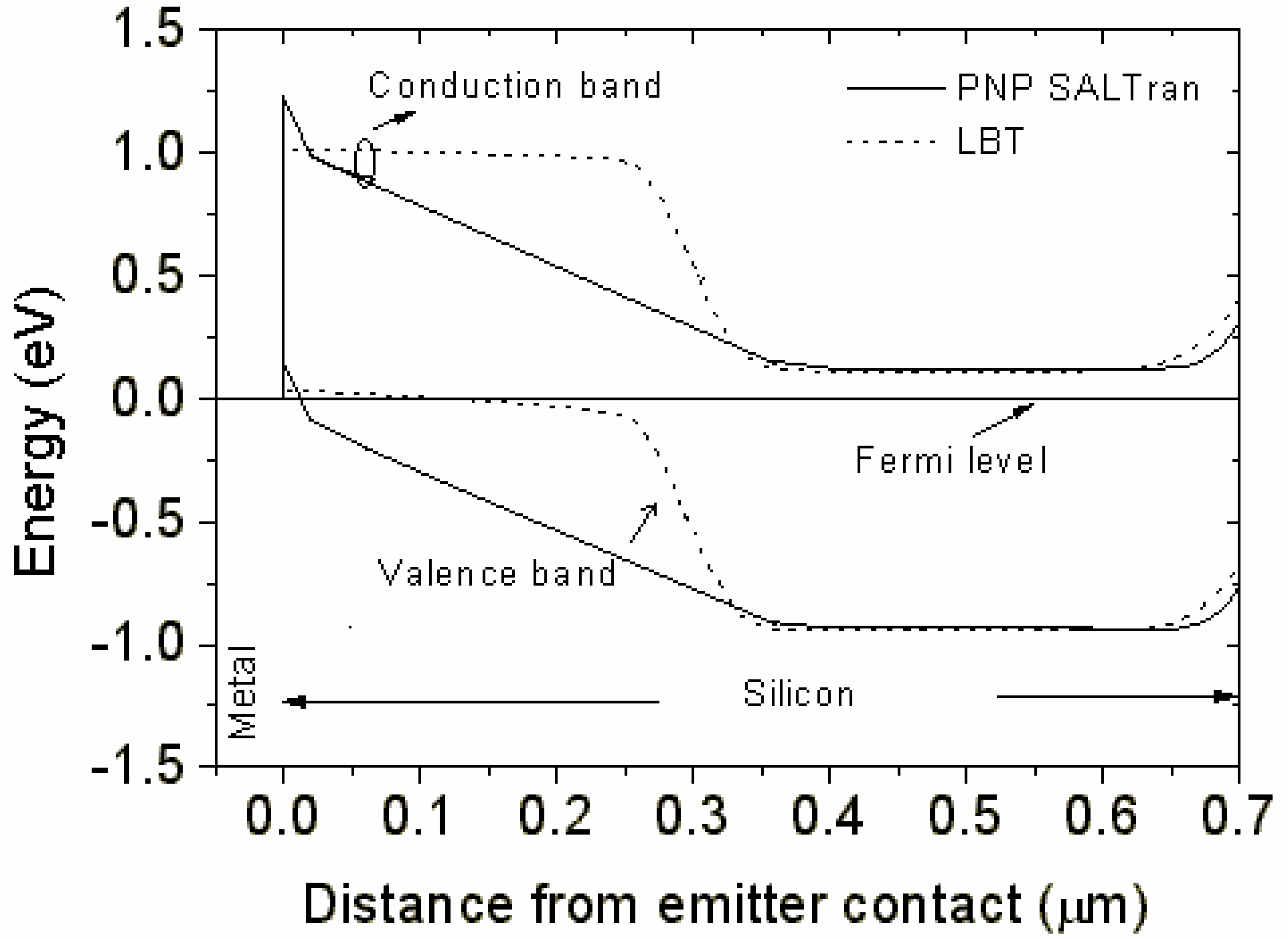

(a)

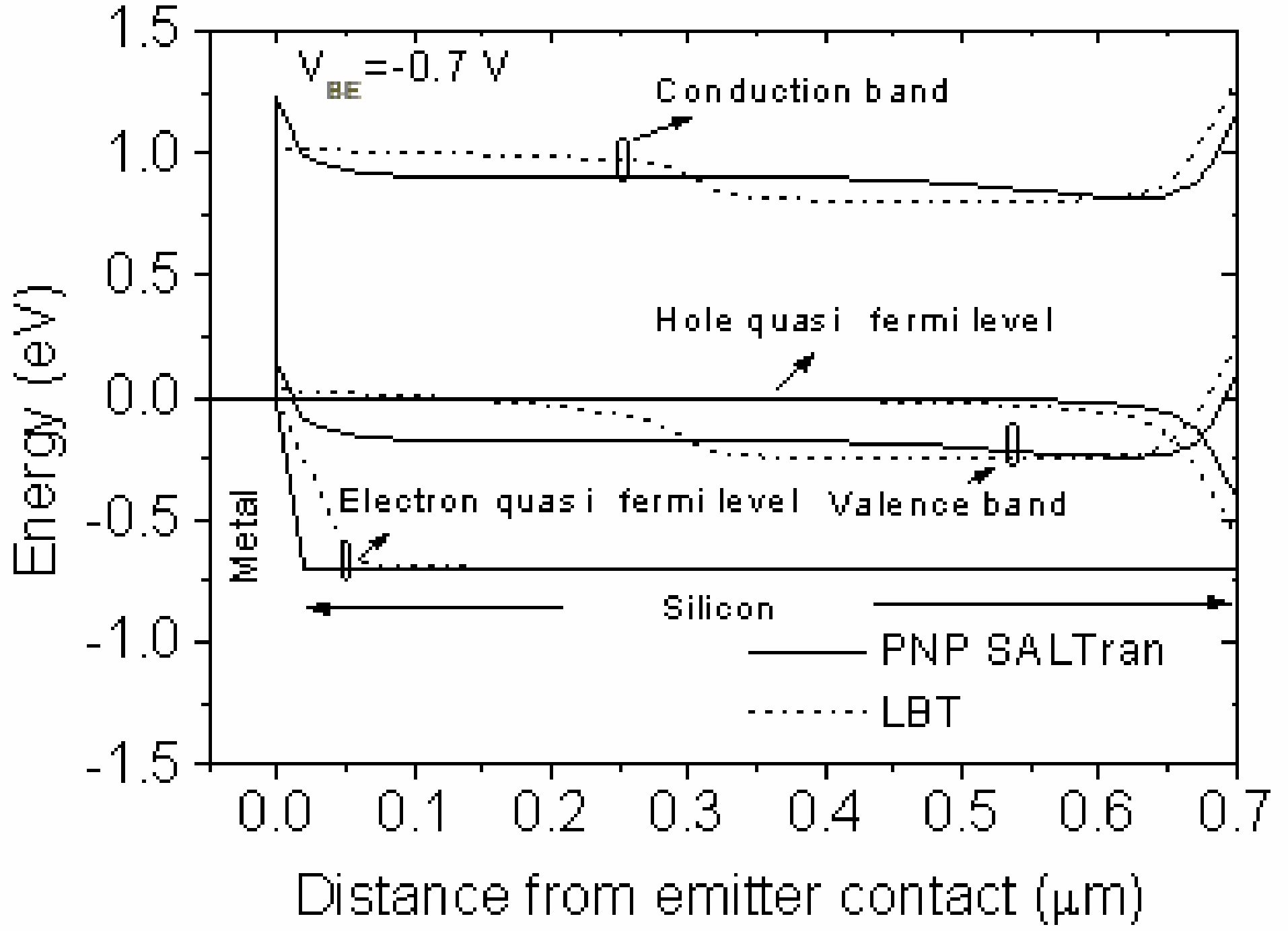
V BE =-0.7 V
Conduction band
Hole quasi fermi level
Electron quasi fermi level
Valence band
Metal
Silicon
PNP SALTran
LBT
Energy (eV)
Distance from emitter contact (μm)

(b)

Fig. 4

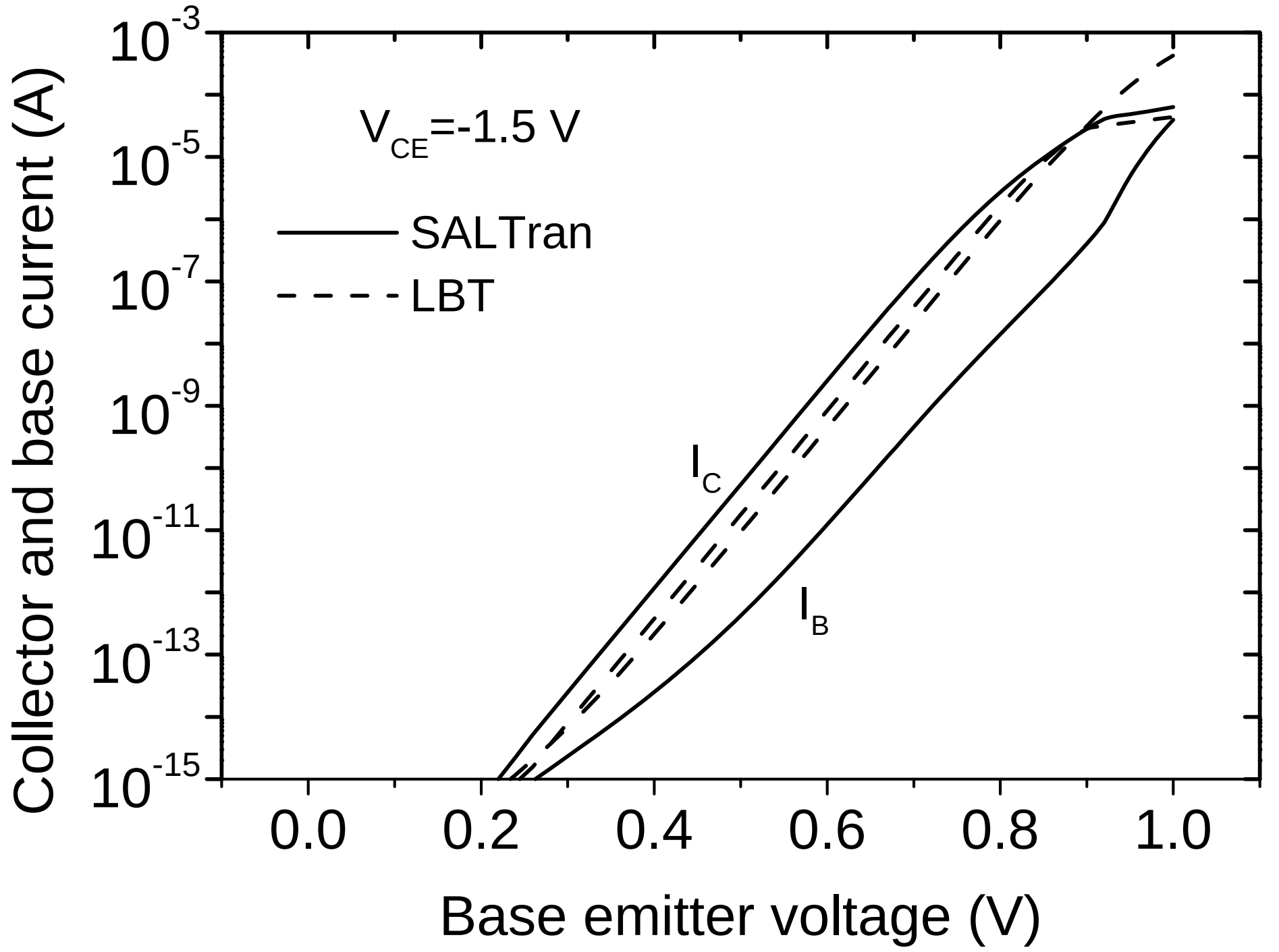

(a)

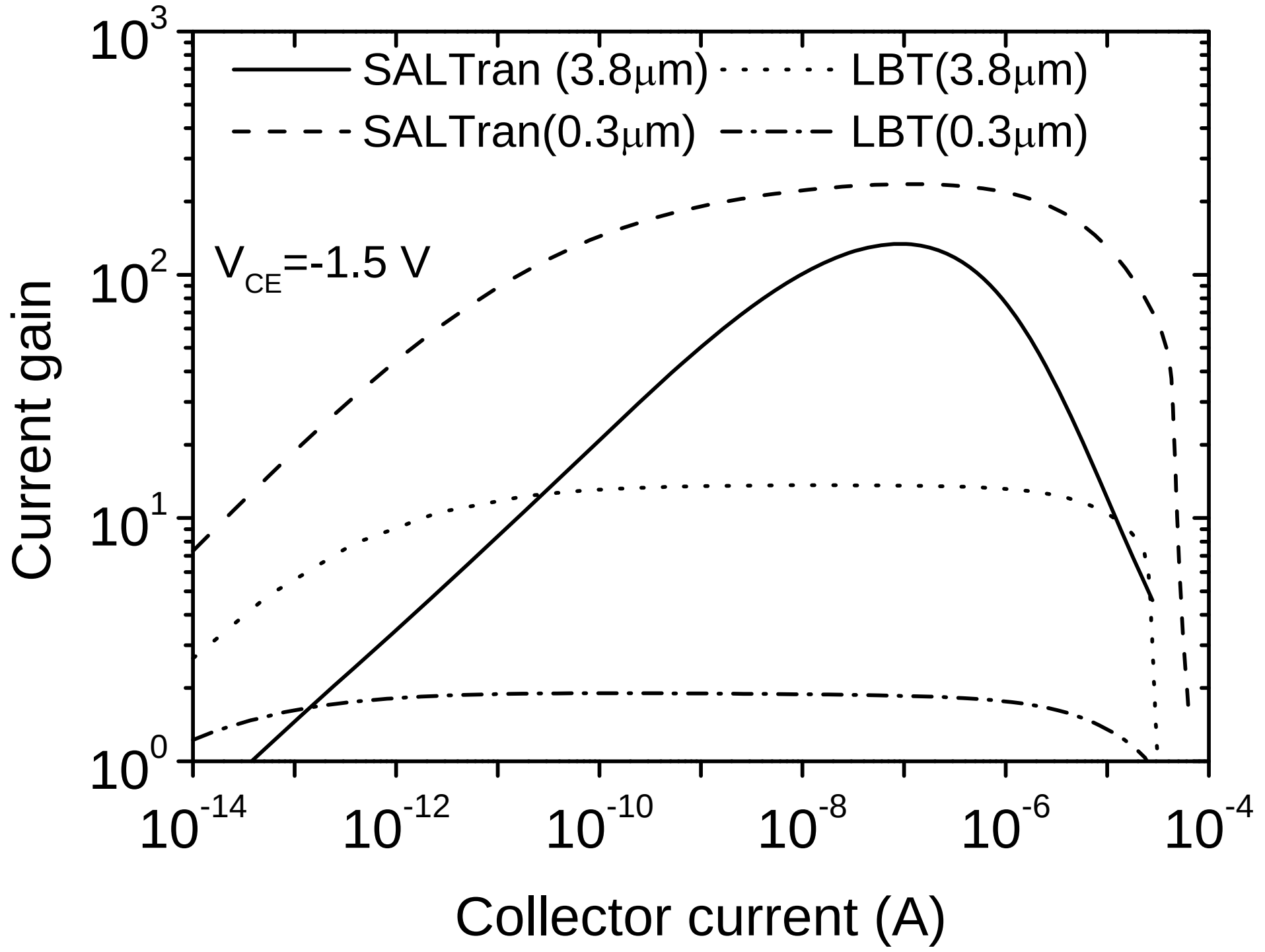

(b)

Fig. 5

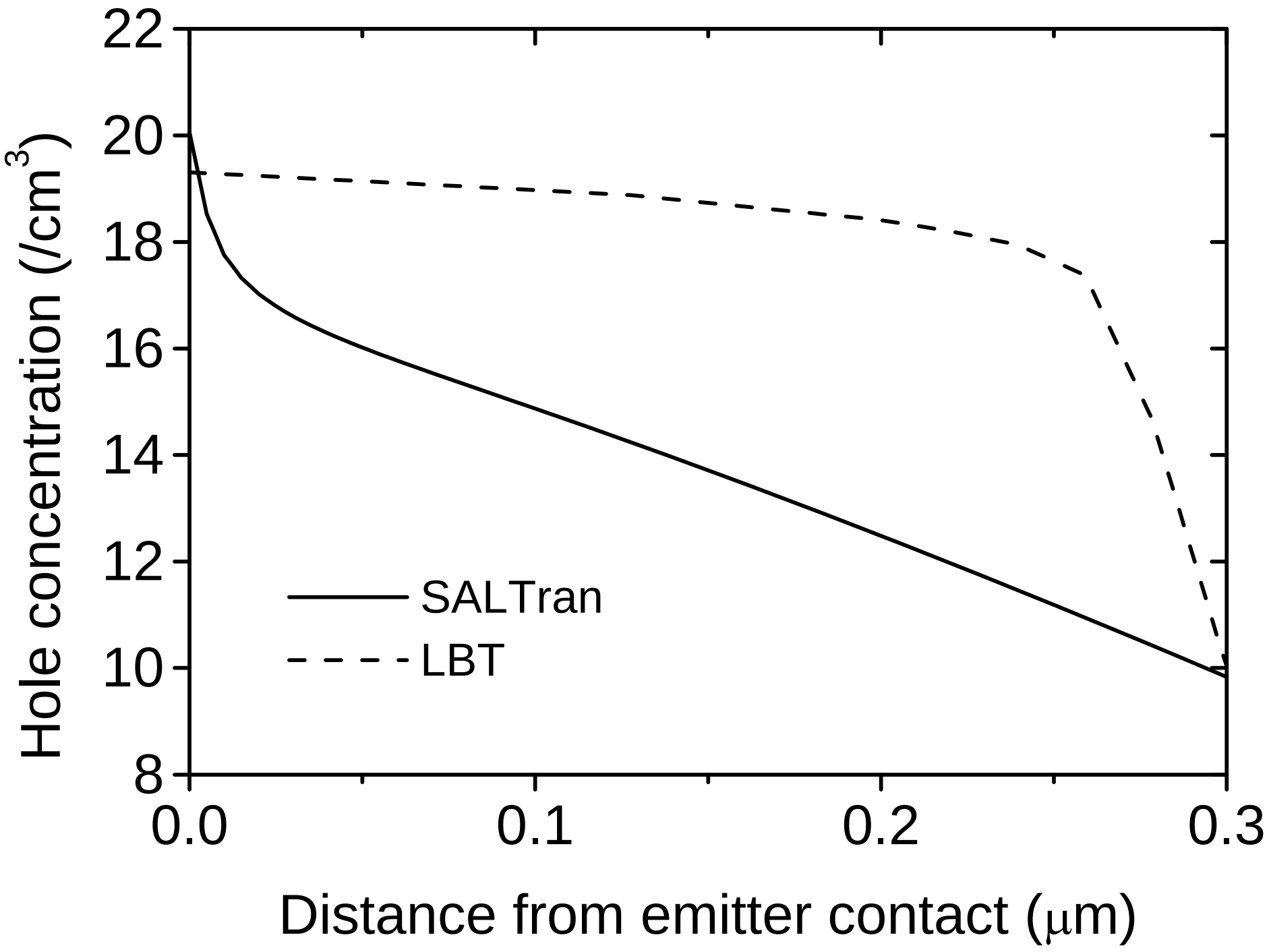

(a)

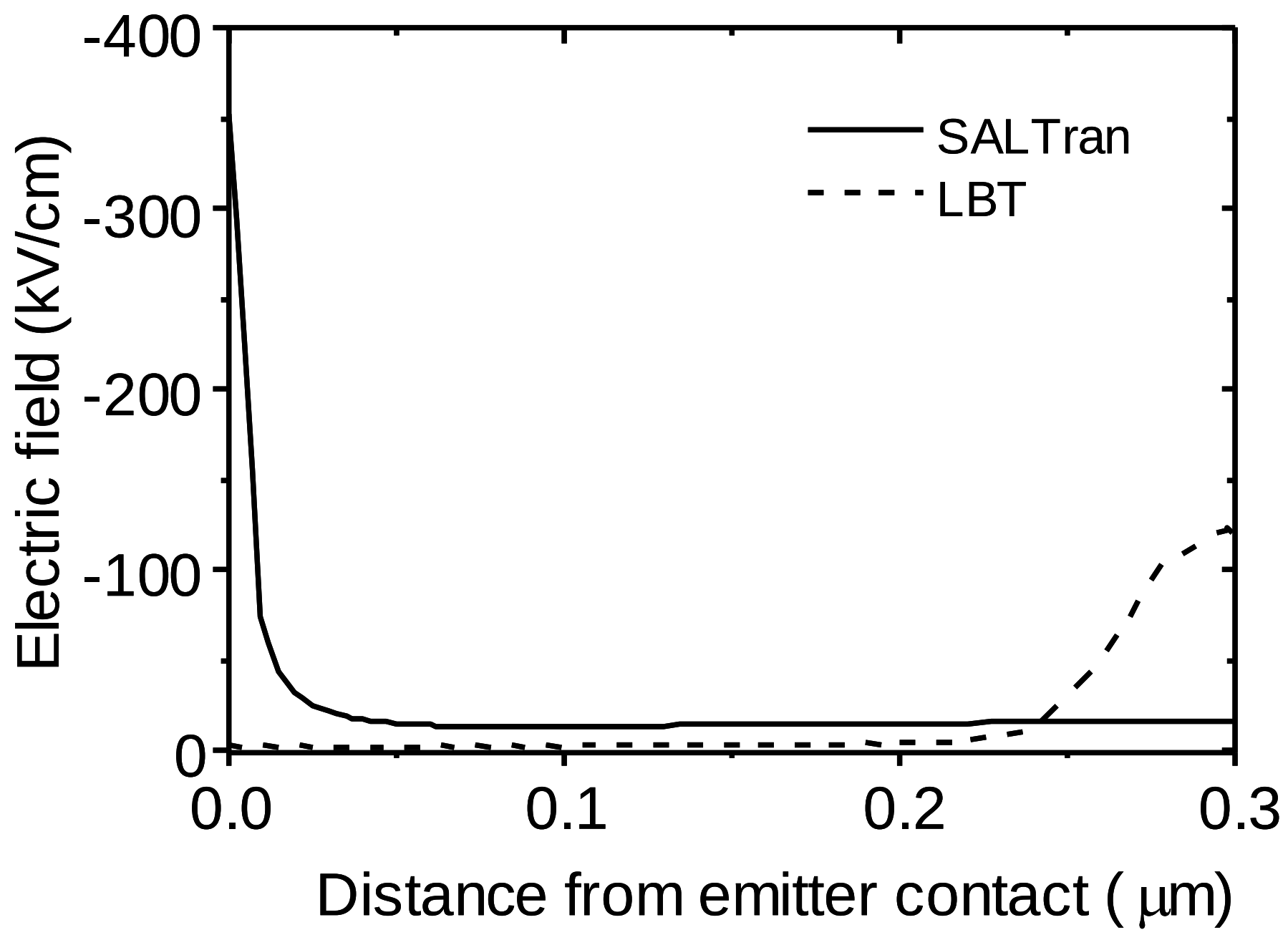

(b)

Fig.6

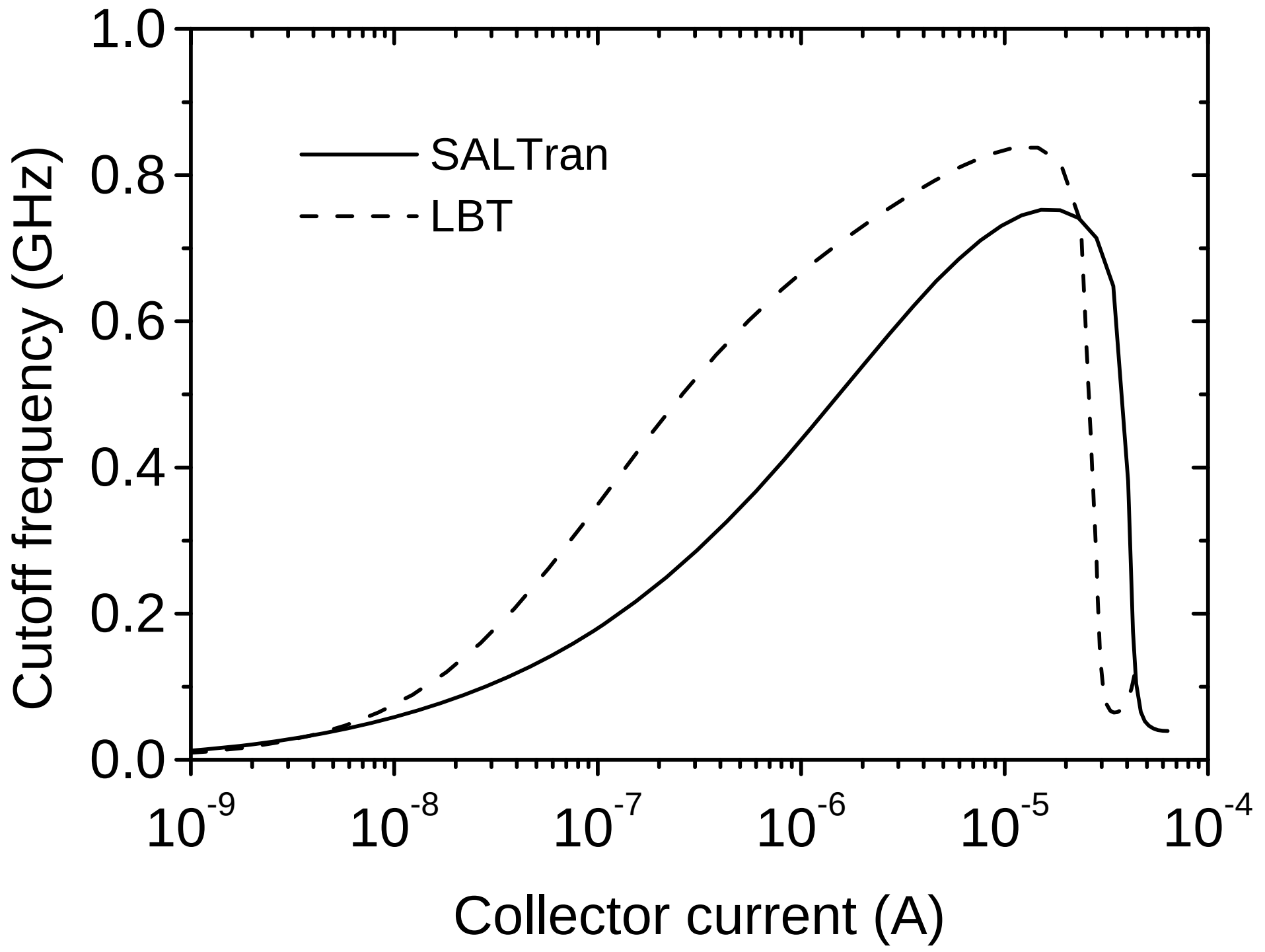

Fig. 7

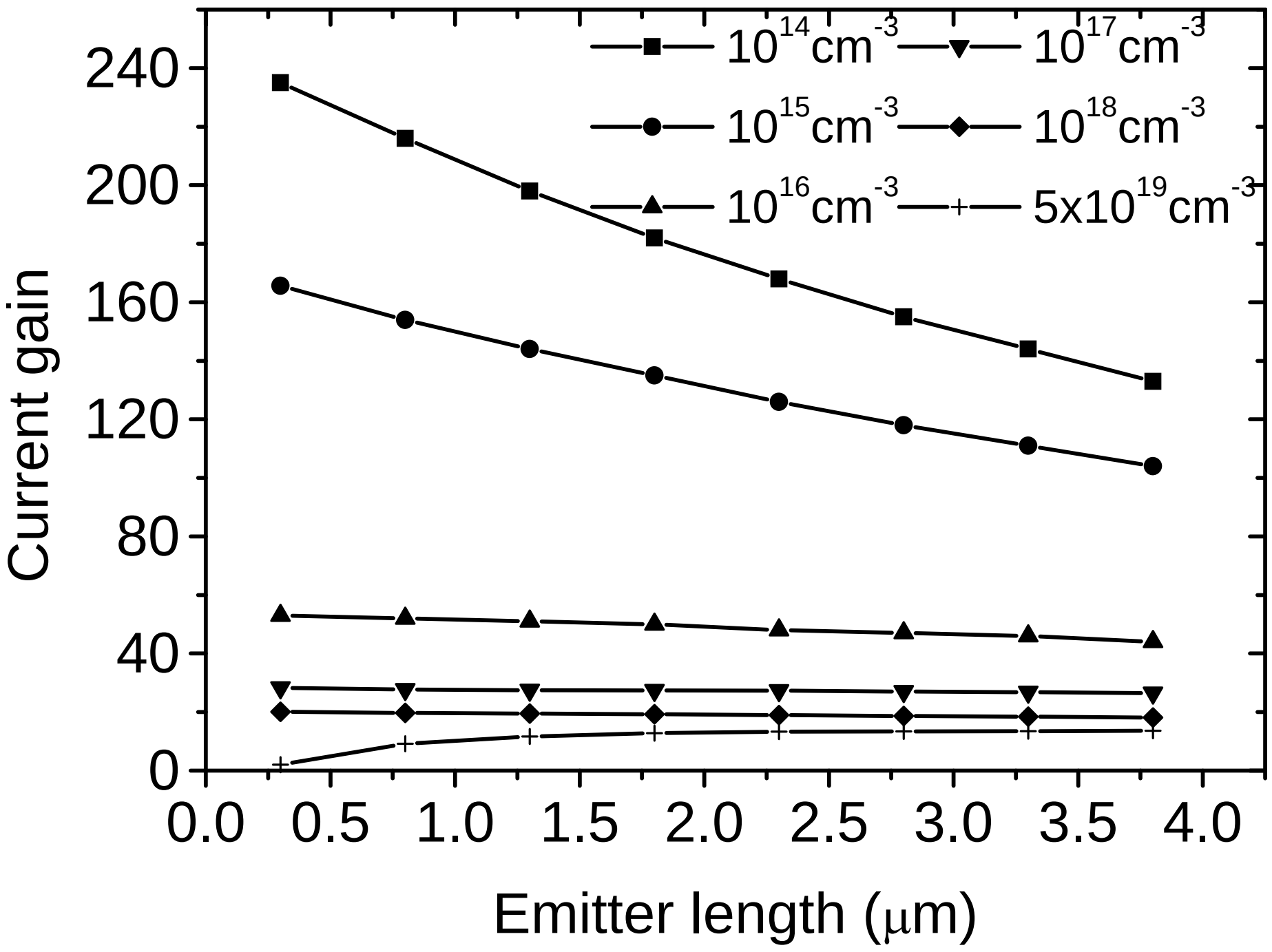

**Fig. 8**

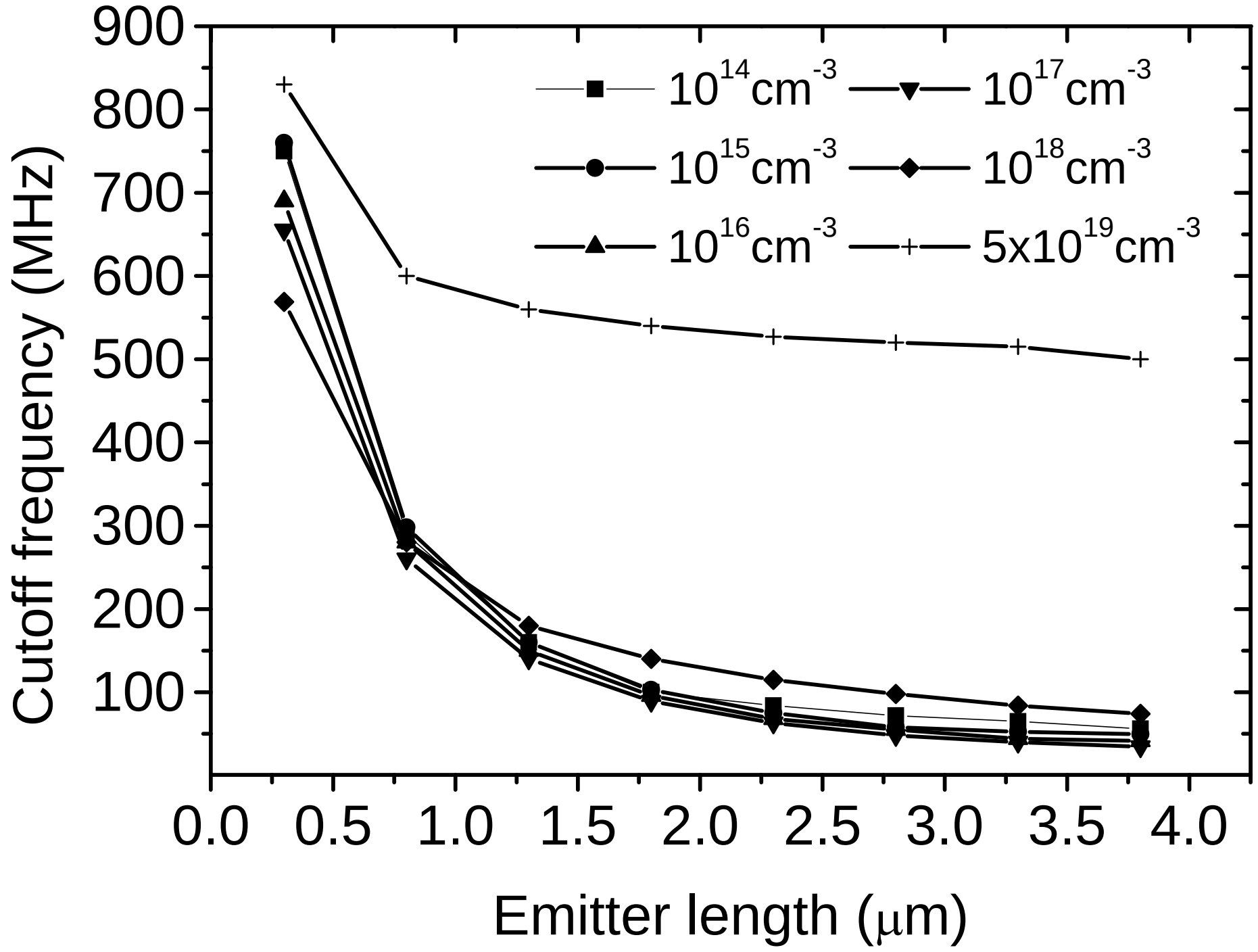

900
800
700
600
500
400
300
200
100
Cutoff frequency (MHz)
0.0 0.5 1.0 1.5 2.0 2.5 3.0 3.5 4.0
Emitter length (μm)
10^14 cm^-3
10^15 cm^-3
10^16 cm^-3
10^17 cm^-3
10^18 cm^-3
5x10^19 cm^-3

Fig. 9